# Architecture of a massive parallel processing nano brain operating 100 billion molecular neurons simultaneously

Anirban Bandyopadhyay[1]*, Daisuke Fujita[2], Ranjit Pati[3]

*[1]*Advanced Scanning Probe Microscopy Group, Advanced Nano Characterization Center, [2]*Materials and Nanoarchitectonics (MANA), National Institute for Materials Science 1-2-1 Sengen, Tsukuba, Ibaraki, 305-0047 Japan. Department of Physics, Michigan Technological University, 1400 Townsend Drive Houghton, MI 49931 - 1295, USA*

**Molecular machines (MM, Badjic, 2004; Collier, 2000; Jian & Tour, 2003; Koumura & Ferringa, 1999; Ding & Seeman 2006) may resolve three distinct bottlenecks of scientific advancement. Nanofactories (Phoenix, 2003) composed of MM may produce atomically perfect products spending negligible amount of energy (Hess, 2004) thus alleviating the energy crisis. Computers made by MM operating thousands of bits at a time may match biological processors mimicking creativity and intelligence (Hall, 2007), thus far considered as the prerogative of nature. State-of-the-art brain surgeries are not yet fatal-less, MMs guided by a nano-brain may execute perfect bloodless surgery (Freitas, 2005). Even though all three bottlenecks converge to a single necessity of nano-brain, futurists and molecular engineers have remained silent on this issue. Our recent invention of 16 bit parallel processor is a first step in this direction (Bandyopadhyay, 2008). However, the device operates inside ultra-high vacuum chamber. For practical application, one needs to design a 3 D standalone architecture. Here, we identify the minimum nano-brain functions for practical applications and try to increase the size from 2 nm to 20 μm. To realize this, three major changes are made. First, central control unit (CCU) and external execution units (EU) are modified so that they process information independently,**

**second, CCU instructs EU the basic rules of information processing; third, once rules are set CCU does not hinder EU-computation. The basic design of the proposed nano-brain is a dendrimer (Hawker, 2005; Galliot, 1997; Devadoss, 2001; Quintana, 2002; Peer, 2007), with a control unit at its core and a molecular cellular neural network (m-CNN, Rosca, 1993; Chua, 2005) or Cellular Automata (CA, Wolfram, 1983) on its outer surface (EU). Each CNN/CA cell mimics the functionality of neurons by processing multiple bits reversibly (Rozenberg, 2004; Li, 2004; Bandyopadhyay, 2004). We have designed a megamer (Tomalia, 2005) consisting of dendrimer (~10 nm) as its unit CNN cell for building the giant 100 billion neuron based nano brain architecture. An important spontaneous control from 10 nm to 20 μm is achieved by an unique potential distribution following $r = a\sin k\theta$, where *r* is the co-ordinate of doped neuron cluster, *k* is the branch number, *θ* is the angle of deviation and *a* is a constant typical of the megamer architecture.**

# I. Introduction: essential functions of nano-brain

The core architecture of a nano-brain whether operating in a nano-factory, functioning as a nano-surgeon or nano-computer could be the same. The architecture may consist of a control unit connected to all execution parts following *one-to-many* communication principle. In our recently described proto-nano-brain (Bandyopadhyay, 2008), we have demonstrated this principle in practice. The principle states that, if a large number of molecules are connected radially to a single molecule then by tweaking the central molecule one can logically control all radially connected units at a time. To control the logic operation of a large assembly, we need to control only the central molecule, which we name the central control unit (CCU). Currently, the CCU can send only one instruction without external interference. We wish to develop it in such a way that it is able to send a series of logical instructions to the execution units (EU) during its operation. Only then, the complete architecture would execute series of operations one after another by itself, independent of any external stimuli or human interference (Koumura, 1999). This is important as it is not practical to instruct the control unit of a nano-factory several times for completing the task, or instruct the control unit of a nano-computer at every stages of its derivation of a math problem, or advise the control unit of a nano-surgeon its next move during a brain operation.

The fundamental element that constitutes a molecular nano-brain is a molecular neuron. A neuron is an analogue switch. Beyond a threshold voltage, continuous increase of applied bias should generate more than two conducting states in a neuron-like molecule. Unfortunately, almost all reported practical single molecule switches are binary (Chen, 1999). We reported the first 2 bit single molecule switch operating reversibly between four conducting states (Bandyopadhyay, Miki 2006). Since then we have tested several multi-electron processing organic systems and invented as high as 4 bit molecular switch operating reversibly between 16 distinct conducting states (Simic Glavasky, 1989). Compared to an analogue switch, 16 choices may appear low, however, because of its simultaneous response in coherence with all neighbors, it can process massive information compared to the existing processors. Multi-level switches are essential to build nano-brain because of two fundamental reasons.

The first rational originates from the fact that, while designing the hardware of a nano-brain, we adopt one major paradigm shift for information processing and acquiring the output data. Information is processed horizontally through the surface, and it may be a euclidian or a non-euclidian surface. However, the output is always accessed vertically to this surface. If information processing and information acquisition directions are the same then signal processing noise is the product of all noise functions generated at every step on the processing path (Figure 1 a). For a vertical observer, every action executed on the planar surface is mapped directly, therefore total output noise is simply the noise function generated at one particular node. Now, a versatile decision-making machine could be realized in the conventional in-plane processing circuits by measuring output at different spatial locations on a surface. For orthogonal processing and acquisition, the only possibility for versatile decision-making is to increase the number of logic levels.

To create an adaptable hardware, multi-level switches are essential. Adaptation within organic neural matter means that it adopts particular global rules for the decision-making process and all local rules are played within the framework of those global rules (Kohonen, 1989). On the organic monolayer or in a molecular assembly if we wish to mimic this particular feature, then it would be needed to create a background logic map. This map would define the minimum energy surface for the output pattern and all further solution sets generated on this surface would have higher energy than the background logic map (Figure 1 b).

Self-organisation is essential for a learning hardware, by which processing units re-arrange themselves in such a way that the input signal is re-distributed among processing units in a new pattern. Instead of directly processing the receiving signal, bio-processors adopt self-organisation to convert it into a signal essentially acceptable to the hardware. An integrated molecular assembly undergoes a global relaxation process to reach a new energy minimum (Figure 1 c). If structural relaxation is correlated with spontaneous logical modification of the received signal, then self-organization process could be tuned within the hardware. This is important as one can specifically identify parameters to encode biological *learning process* into an electronic hardware. Self-organization does not necessarily mean that the processing units would physically re-assemble themselves into a different conformation/arrangement; rather it could preferably be reconstruction of a logical map.

As discussed above, optimally efficient *one-to-many* communication may require a spherical architecture. The reason is that $360^o$ rotation of our disk-shaped 2 D nano-brain (Bandyopadhyay, 2008) around its planar axis is a sphere (Figure 1 d). The basic information-processing unit of such spherical assembly is a molecular neuron, essentially a single molecule multilevel switch. Three basic criteria are suggested to design processing circuit of genuine unconventional computers (Dewdney, 1984; Benenson, 2001; Păun, G., 2000). First, there should be a wireless network of elementary processing units as heating is one of the most important problem for the future generation processor (a Pentium IV processor has ~10 km wiring/cm$^2$). Second, the processor should be massively parallel capable of computing more than several hundred bits at a time. Third, it should spontaneously evolve logical decisions or execute emergent computation. One example for such computation is information processing through a cellular neural network (CNN, Rosca, 1993; Chua, 2005; Wolfram, 1983). One possible way to realize a CNN would perhaps be to identify governing equation or cellular automata (CA) rules of an organic monolayer composed of molecular neurons. If well-defined CA rules exist that enables the monolayer performing meaningful computation, we can say that the molecular neuron functions as a CNN or CA cell. Then by assembling these molecular neurons we can construct a euclidian spherical information processing surface instead of a planar CA. All molecular neurons on this surface should be connected to the central core. The central core should function as CCU and wired architecture would define the global relaxation function controlling self-organisation and adaptation, which in turn would control the learning process.

# II. Basic design principles

## A. Molecular neurons

Recently we have shown that molecules with particular design can exchange one or more electrons and undergo reversible phase transition by changing the structural symmetry. Cyclic voltammogram (CV, Chambrier, 2006) can map these reversible phase transitions and one can measure quantitatively the number of electrons spent to induce these phase transitions into a molecule. By studying a typical Fe-terpyridine linker molecule Pati *et al* have shown that change in orbital symmetry and phase transition leads to *negative differential resistance* (NDR) peaks (Pati, 2008). The *2,3,5,6-tetramethyl-1-4-benzoquinone* (DRQ) molecule exhibits two NDR peaks in its current voltage (*IV*) spectrum (Figure 2 a). Using these two NDR peaks, we have demonstrated reversible 2 bit information processing (Bandyopadhyay, 2003). Simulations show that this molecule undergoes phase transitions twice during exchange of one and two electrons reversibly. Theoretically, if we find a molecule that undergoes multiple distinct phase transitions by changing its symmetry and experimentally the same molecule shows multiple reversible peaks in CV, then it could be a suitable candidate for single molecule multilevel switch (Figure 2 b). These particular kinds of molecules should exhibit multiple NDR peaks in the *IV* characteristics as current across the molecule would increase as soon as a particular phase transition takes place and then current would decrease leading to a distinct relaxed phase prior to another phase transition. To confirm that such a large number of NDR peaks are not noise, rather a molecular response we test *random access memory* (RAM) operations with millions of consecutive *write-read-erase-read* pulses (Bandyopadhyay, 2003).

## B. Step-function molecules or Q-dots

A multiple NDR effect in a single molecule mimics electronic response of a circuit where similar numbers of *resonant tunneling diodes* (RTD, Potter, 1988) are connected in series. However, CNN theorists have proposed a few alternate *IV* characteristics of CA cells. One of them is saturated bi-polar synaptic feature $I(V) = 1/2(|V+T|-|V-T|)$, see figure 2 (Chua, 2005). As no literature exist on the practical realization of such electronic behavior using a single device, here we propose a new methodology. The step function has two basic characteristics. First, the device should show *conductance at zero bias* (ZBC, Liang, 2002). ZBC is realized only when a device demonstrates Kondo effect. The Kondo effect is normally realized in magnetic materials at very low temperatures. In single molecules, the Kondo effect could perhaps be realized even in room temperature. Q-dots localize spin density; therefore, an artificial molecule composed of Q dots is a potential alternative to the specially designed molecules (Cronenwett, 1998).

Apart from ZBC, the single molecule or Q dot would require one additional feature in its *IV* characteristics. Output current should remain independent of external bias or remain constant outside the ZBC region. This is hardly possible to realize ideally in a practical device since, higher electric field tend to inject more carriers into the device. Statistically several reported devices reporting the Kondo effect show that within a particular range beyond ZBC, output current remains constant (Temirov, 2008). The constant current output beyond ZBC region may be a natural phenomenon in these devices; since ZBC is induced by scattering of carriers with the localized spin part of a system and a neutral region exists both sides of the ZBC (outside ΔV region in Figure 2 c) before the onset of the natural tunneling regime. These features, ZBC and nearly constant currents at both sides of the ZBC peak are observed in a particular conformer of Rose Bengal molecule. Our resent studies on Rose Bengal derivatives (Bandyopadhyay, submitted, Figure 2 d) show that, particular conformer of this molecule exhibit this particular step function like *IV* characteristic in room temperature.

Three other synaptic features which could be used as CNN cells are respectively linear bi-polar which is devised as a resistor, a rectifier (Chua, 2005) and a memristor or variable resistor (Figure 2 e, Strukov, 2008; Chua, 1971). Single molecule analogue of a variable resistor could be realized in a conducting polymer chain (Terada, 2000), and donor-acceptor group coupled single molecules may be used for operating as a molecular rectifier (Metzger, 2000). The single molecule memristor is not invented yet, though several thin film devices are reported in the literatures that demonstrate similar performance (McGinness, 1974; Potember, 1979; Mukherjee, 2005).

## C. Cellular Neural Net of molecular neurons

Organic molecules functioning as neurons or exhibiting step-function in the *IV* spectrum could be used to build *cellular neural net* or CNN. The basic criterion in order to use the molecular neuron as CNN cell is that, in the molecular assembly, all neurons should retain their reversible switching between multiple conducting states. In addition, once a set of logic states are written by applying electrical bias to a set of molecules, the written logic pattern should change reversibly following particular rules characteristic of the molecular assembly. The conventional approach to build molecular assembly is to grow a monolayer on an atomic flat metal surface by e-beam evaporation of molecules in an ultra-high vacuum (UHV, $10^{-8}$ Torr) chamber, heating the molecule above melting point. Another approach is dropping micro/nano molar molecular solution on a fresh atomic flat substrate. The monolayer is deposited on the atomic flat surface such that the logic state distribution pattern is read accurately by measuring a tunneling image using *scanning tunneling microscope* (STM). Analyzing a series of images, we identify common rules that govern transport of logic states on the surface, change in patterns and spontaneous creation/destruction particular logic states. We name these rules as CA rules and the literature is rich in analyzing the methods of computation using CA rules (Figure 3 a, Wolfram, 2002).

We have deposited bilayer of a 2-bit switching molecule, *2,3-Dichloro-5,6-dicyano-1,4-benzoquinone* (DDQ) on the Au (111) surface in an UHV chamber in particular conformation. Note that, for molecular neurons, it may be useful to examine crystal structures of the parent and related molecules and symmetry of the deposited monolayer, since conformer associated with different conducting states may give rise to structurally distinct monolayers, following distinct CA rules. Ordered molecules in a well-defined symmetry mimic a classical mathematical model of CNN. By imaging the surface continuously for hours we have determined CNN rules for DDQ (-1) monolayer on the Au (111) surface (Figure 3 b).

A classical organic monolayer is a CNN compatible monolayer if particular features are encoded. These features are as follows. First, the electron-density distribution pattern (or logic map) evolve consistently on the surface, it should not spread out completely over the monolayer and not even localize disappearing permanently from the surface. In both cases, CA rules do not allow versatile computation. Second, logic states should preferably depend on the number of electrons trapped in the molecule not on its conformational changes. Then information processing would be re-distribution of electrons. Since in the monolayer conformational changes are probabilistic event, it might cause unwanted errors during computation, -invalidating the CA rules. Third, packing of molecules should be such that following a sequence of changes in orientation a cluster of molecules would return to the same state. This particular phenomenon allows a few neighboring molecules to execute unique local CA rules and allows the system to respond to global changes at a much faster rate, for instance deleting entire pattern from the CNN surface.

To demonstrate a practical application, we have created monolayer of Rose Bengal (RB) molecule in which the electronic property of the particular RB conformer that demonstrates step function in the *IV* measurement is pronounced. The only possibility for such a response is that in the confined state, the molecule can switch only to a single conformer state. Therefore, the surface could serve as a potential CNN template (Figure 3 c, d).

Irrespective of the electronic response of a nano-scale CNN cell, be it a constant resistor, variable resistor (memristor), rectifier, or multilevel NDR switch, when packed in the form of a molecular assembly, the integrated system should function as a cellular neural network with distinct CA rules. Note that the difference between CNN and classical CA is that, a CNN follows a particular equation to determine the next state of a cell whereas CA has a particular set of rules to do the same. Here, the proposed CNN surface can have any of these features.

## D. Spherical assembly

To create a standalone processor, we propose a practical CNN requires to be constructed on the surface of a stable spherical assembly. One option is to use dendrimers (Galliot, 1997; Devadoss, 2001; Quintana, 2002) or dendritic molecules (Hawker, 2005; Peer, 2007) as the core architecture and replace the end groups with molecular neurons. Alternatively, molecular neurons may be anchored covalently/non-covalently with the end-groups. The number of end-groups increases with the higher generations of dendrimer generally to the power of two (Figure 4 a). Chemists can modify the surface of the dendrimer with distinct number of molecular neurons constructing a CNN. For the CNN to operate on the sphere all molecular neurons on the spherical surface should be packed densely in a similar way to the organic monolayer on a planar surface. Otherwise, a universal equation determining the next state of a cell, or the CA rules cannot be determined accurately (see DRQ neurons on G2 PAMAM dendrimer movie, http://in.youtube.com/watch?v=5S8UW_3bxlg , Figure 4 b, c). Therefore, for a particular dimension of molecular neuron, there is a threshold value of dendrimer-generation (say G8 or G9), which gives rise to an atomically packed CNN surface. An ordered atomic pack is essential for the well-defined transport of electrons or logic states through the surface.

------------------------------------------------------------

The CA rules or a universal governing equation for the CNN on the spherical surface would be different from the rules/equation of CNN created by same neurons on the planar surface. Dynamics of the dendritic branches would determine global relaxation of molecular assembly. Therefore, magnitude of electric bias for reading and writing of bits on the molecular neurons assembled on the dendrimer surface would be different from the monolayer constructed on the atomic flat metallic/semi-conducting surface. There are two major factors determining the CA rules or CNN governing equation on a spherical CNN surface.

The most important factor is orbital conjugation of the dendritic branches in the spherical core (Devadoss, 2001). If the core is conducting and forms polarons as electric pulse is applied to write a particular state on the CNN surface, polaron mediated conduction throughout the core may modify the logic pattern written previously. The second factor is the quadrupole moment (Figure 4 d). The extremely dynamic architecture does not remain ideally a sphere during operation. Therefore, the quadrupole moment of the system may play a major role in deriving final solution pattern on the CNN surface.

We have deposited non-conjugated PAMAM dendrimers on Au (111) surface (Galliot, 1997; Devadoss, 2001; Quintana, 2002). Fixing the STM tip on a single dendrimer ball we have sent multiple random pulses on 2, 5, 6, 7 and 8 generation PAMAM dendrimers with amine and COOH end-groups and analysed the output current as if the complete system is a single unit (Figure 4 e). Continuous RAM operation detected more than 80 conducting states in each system suggesting that even in the non-conjugated systems symmetry induced quadrupole moment enables generation of distinct conducting states of the system. Until now, we are not able to map the logic pattern distinctly on the surface, which would be a statistical image since only the upper hemisphere is visible to the STM scanning (Figure 4 e). To resolve the problem we are presently placing the dendrimer ball into a nano-gap device (Khondaker, 2002) and applying horizontal electric bias across the dendrimer ball while conducting states are probed vertically from the top surface. This technique not only helps to identify the pristine conducting states of the dendrimer, an additional gating effect helps to resolve the sub-conducting levels apparently not visible in the two probe measurements. Calculation further shows that these sub-conducting levels are strongly correlated to its relaxation symmetry. Simultaneous three-electrode measurement (Bandyopadhyay, Nittoh, 2006) of conductance variation is more reliable than conventional two terminal measurements for mapping the dynamic relaxation of a nano-brain.

Dendrimer-like spherical assemblies may trap several electrons at a time, since a sphere has only one symmetry (Figure 4 f; Balzani, 1998; Nielsen, 2000; Takada, 1997; Denti, 1992). Any molecular assembly with one symmetry cannot sustain distinct patterns of different energies on the outer surface. At a very particular reduction/oxidation bias, all neurons would switch simultaneously to a fixed conducting/logic state without any control as a large number of electrons probably equal to the number of neurons on the surface would be trapped into the system or move out of the system at a time. Therefore, a perfect symmetry may be undesirable for constructing a nano brain. The more possible it is for symmetry induced distinct phase transition of the dendritic architecture, the more versatile would be the information processing. Multi-state electron processing could be recorded using CV similar to the study of reversibility of multiple conducting states in a single molecule neuron. However, simply increasing asymmetry in the architecture would invite another important problem. Externally, encoding a logic pattern on a spherical surface would be most reliable if the surface is perfectly spherical. The reason is that prior to real testing of CNN performance, the governing equation for CNN information processing or the CA rules require to be revealed. Only way to do that is to cover the surface with vertical electrodes fixed at tunneling distances apart. Therefore, a consistent number of symmetry is essential to map evolution of logic states.

## E. The central control unit (CCU) or Core of the sphere

Ideally, any little change in potential or conformation at the central region would generate massive changes in logic pattern throughout the surface. As every neuron on the sphere is connected directly to the center, we are trying out suitable design changes to accommodate couple of switches in a particular pattern around the center of the spherical assembly (Figure 5 a). A key principle to the CCU coding pattern is the modulation of polaron transport across the dendritic conjugated branches inside the sphere. If switching molecules are trapped at the center of the sphere then the polaron transport length is maximum, equal to the radius of the sphere (Figure 5 b).

If complete logic pattern on the spherical CNN surface are required to be modulated at once then a multilevel switch has to be trapped at the very center of the sphere. Depending on the required modulation area of the CNN surface, molecular switches should be trapped at particular branching locations (Miklis, 1997). The smaller the surface area to be modulated, the further is the trapping location from the sphere center. Since conductance level, logic state and number of electrons in the molecule are closely associated, therefore, the 3 D potential map calculated from the molecular switch co-ordinates on the dendrimer branches would be the CCU code (Figure 5 a). The relaxation dynamics of the CCU region would continuously set new kinds of patterns on the CNN surface one after another as shown in the Figure 5 a series. Each pattern would correspond to particular set of CA rules or a CNN governing decisive equation. Therefore, CCU coding enables spherical assemblies to carry out series of decision-making computations one after another, spontaneously.

For computation, we need to write an input pattern on the CNN surface (Figure 5 c). During encoding the input pattern on CNN surface, CCU code might change because of the pulse applied. Note that the central molecule of the sphere would require maximum bias, and the molecules located at higher concentric radii from the center would require lower bias for encoding a conducting state. Therefore, two distinct patterns, one in the CCU and another on the CNN could be encoded. For better accuracy in pattern encoding, different kinds of molecular switches could be used as structurally distinct molecular switches have distinct threshold biases (Figure 5 a).

# III. Operational Algorithm
## A. Dynamics of cellular automata CNN and CCU

It is not possible to analyze dynamics of such a large architecture using fundamental quantum mechanical analysis, as an astronomically powerful computer would be required. Therefore, we follow an alternate method of analyzing the CNN dynamics in terms of CA dynamics known as *basin of attraction* (BA, Wuensche, 1991; Martin, 1984; Pitsianis, 1989). Until now, we are not able to make very particular changes in the mathematical model of BA dynamics formulation so that it isolates the structural and electronic relaxation of CNN hardware. Since the BA dynamics formulation maps the time evolution of CNN logic pattern's distinction at an identity space, comparison of this map with the monolayer would provide important differences caused by spherical shape and CCU dynamics together. However, we need to isolate the role of the spherical shape and CCU from the mixed response so that we can map the effect of the CCU on the global performance of the CNN. One possible way to do that is to construct two spherical assemblies one with CCU and one without it. Comparing BA profile for these two different hardwares we can identify role of the CCU exactly (Figure 5 d). Using statistics, we are trying to develop two sets of grammar rules to decipher the map of these dynamics simulation, one for the CCU and another for the CNN surface.

The BA dynamics of cellular automation is analyzed by identifying a cellular automation transition function. Using this function, all possible states and trajectories are determined and plotted as global behavior of the system. The space-time pattern generated by CA rules would also become a function of CNN pattern and its dynamic relaxation. Two kinds of symmetry one rotational and one bi-lateral survive on the spherical surface. The correlation between structural symmetry of the spherical assembly and symmetrical emergence of CNN's CA rules is an important tool to manipulate continuous relaxation process of the CCU. Its relaxation changes the CA rules operational on the surface CNN. Now, time duration between occurrences of two relaxations requires to be tuned to develop a temporal control on multiple consecutive patterns evolved on the CNN surface (Figure 5 c, 6 a).

## B. Spontaneous relaxation of the sphere

Spontaneous relaxation of the molecular assembly is different from self-organization. The spontaneous relaxation originates from the nature of structural asymmetry and 3 D potential profile of the central core (CCU). While self-organization occurs entirely on the surface, without any change in the potential profile of CCU, spontaneous relaxation restructures the potential profile of CCU and thus redefines CNN governing equation or associated CA rules (Figure 6 a). To understand the spontaneity of the relaxation process, we excite the spherical molecular assembly to a higher temperature or forcefully induce an extra energy to the system, then observe the dynamic relaxation to different local energy levels. The rate of relaxation decreases as the system reaches to a global energy minima (Figure 6 b). Understanding the spontaneous relaxation process is extremely important, as our objective is to run massive parallel processing on the CNN continuously executing series of operations one after another.

Spontaneous generation of new logic pattern in hardware is primary requirement for creativity, but here our concern is not to evaluate degree of spontaneity. Instead, we need to define spontaneity as a predictable event within the stochastic framework. In other words, a global relaxation of the assembly would be part of a continuous relaxation process on the CNN surface. As soon as a global relaxation takes place, local relaxation on CNN surface would continue to take the assembly to a local energy minimum. Finally, CNN relaxation would induce global relaxation of the complete assembly (Figure 6 a). This is necessary, as we cannot allow that time difference changes between two consecutive relaxations. E. Behrman's brain model (Behrman, 2006), driven by Hameroff and Penrose's proposed quantum computation, also suggest stable local minima similar to our global minima, following quantum mechanical fluctuations.

Time delay of the order of seconds is required between two global relaxations for practical applications. This is achieved by connecting the two global relaxations with a series of essential local relaxation processes. Since a number of multilevel switches can make up the CCU, we convert the 3 D map of logic states into an energy map, and study its relaxation process. Conformational change is essentially a slower process and the process becomes even slower with the increase of design complexity and the assembly volume. Higher is the dendrimer generation, lower is the self-diffusion of atoms in the dendritic chain (see Fricks law, Frenkel 1996). Mean square deviation of atoms remains nearly constant at higher than 6 generation. Therefore, volume distribution of potential is an important factor. Structural relaxation studies have further shown that an isolated large negative potential region at different parts inside the sphere redistributes logic states in such a way that effectively the total electric field is also minimized. Moreover, global relaxation is initiated by a very particular design where two differently charged lobes are connected by a small channel (Figure 6 c). Therefore, using this particular design trick, we can tune the number of global relaxation we require. In addition, the time between two relaxations is entirely determined by density of charge distribution inside the sphere.

## C. Self-organization and learning input pattern

As soon as a distinct pattern is written on the surface of the spherical molecular assembly, the pattern undergoes certain changes and the architecture stabilizes generating a modified logic pattern on the CNN. However, our preliminary study on the assembly shows existence of particular changes in the logical pattern evolution, with the kind of logical operation performed repeatedly on the CNN surface (Figure 1 b). For the hardware, this is hysteresis, and for software, i.e. the logic pattern, this is learning. The system is called adaptive.

Hysteresis driven by local environment of the structure initiates certain changes so that particular rules are favored in comparison to others. When new kinds of patterns are processed repeatedly on the CNN surface, previous hysteresis is removed and a new hysteresis is generated in the system. In other words, one kind of learning is erased and a new knowledge is learnt (Figure 1 b bottom). We feed multiple sets of logical input and output patterns into neural network package analyzing learning process in terms of *adaptive linear element*, *perceptron* and *learning matrices* formalisms (by determining f(L), Kohonen, 1989). Note that, simple step function based devices or memristors are mathematically simpler than multilevel molecular neurons, but no learning process has been detected in the monolayers of memristor molecules in our preliminary studies.

------

## D. Mutually co-existing spontaneity and impulsive operation

In certain cases, nano-brain should execute a complete operation spontaneously and in some cases, it should follow computational instructions blindly. It is already mentioned that a very particular kind of potential distribution enables the assembly to induce a global structural relaxation. If this is entirely a materials property then we cannot perform any computation that requires only one global relaxation of particular kind. Impulsive operation means computation that occurs between two global relaxations (Figure 6 a).

Distribution of multilevel switches inside the spherical assembly is not the absolute condition for encoding a particular sequence of global relaxation, we also need to apply certain large electric pulses to activate CCU potential profile inside the assembly (Figure 5 c). Therefore, we can select particular kind of impulsive computation in the wide bandwidth of energy spectrum, or a set of impulsive computation wherein spontaneous and impulsive computation may co-exist (Figure 6 a, b).

# IV. Coherency of the architecture during operation: factual parametric analysis

## A. Size, dimension relation of a nano-brain

In a nano-brain, computation may emerge at nano-scale but its complete architecture should have a size that can be interfaced using existing nanotube or nanowire electrode based characterization set-ups. Otherwise, nano-brain would face the same fate as molecular (Hipps, 1991) or sub-molecular (Ami, 2002) electronics is facing today. Say, we have an atomically flat crystalline spherical Au (111) ball of 1 $cm^2$ then on top of this surface we can adsorb 100 billion molecules each having 1 $nm^2$ area in a spherical monolayer formation. 100 billion is possible on a sphere of 1 $cm^2$ area if we can interface 1 atomically sharp electrode in 100 $nm^2$ area, thereby 1 billion electrodes around this surface. Using one billion electrodes, one can write one billion bits at a time on this sphere and read same amount of data at a time. Therefore, it is potentially a 1 billion bits parallel processor. Biological cells have dimension of the order of 50 μm. To transport cells in our body, nano-brain carriers should be ~ 1 mm, however, to carry out operation inside a cell, its size should be ~ 10-20 nm or size of a virus. It has to withstand immense dynamic movement of the species inside a living body. This is more valid in a nano-factory where a considerable physical force would be exerted to the nano-brain during operation.

Therefore, if nano-brain seed is sufficient for the need, we may use it as is; however to meet the requirement of complexity we may need to create architectures varying from 20 nm to ~ 50 μm. We can assemble many such nano brain seeds following the same design of nano brain seed as explained in Figure 1 d (right) having various radius, however, here the basic construction unit would be the nano brain seed, with different distinct performances. Undoubtedly, they would provide much better flexibility in operation to the giant nano brain.

Here we concentrate on dendritic nano-brain architectures since they are simplest model to study, however, the *one-to-many* principle could be realized in several unique assemblies, for an example icosahedral virus geometry (Casper, 2004). A dendrimer has total $n_{bc}$ numbers of branches, $n_{bc} = n_c[(n_b^{G+1} - 1)/(n_b - 1)]$, where $n_c$ and $n_b$ are the functionality of core and monomer, $G$ is the generation number. If we plot the number of neurons that could be attached to the surface ($n_t = n_c n_b^G$) with its size ($\propto r^2$, where $r$ is the radius), number of neurons varies particularly with its radius and soon reaches to a singular point. The number of neurons or end groups could be increased until end-groups are atomic distances apart. Beyond that limit (~$10^3$ neurons or 10 generations), no growth is possible. If we wish to assemble 100 billion neurons without using metal core or hollow sphere, on a ~10 nm nano brain seed, $10^3$ concentric spherical layers of dendrimer are required to be fused (considering one nano brain seed volume ~ 350 nm$^3$), see mechanism of such fusion in the Figure 7 a. Final size of the fused megamer brain would be ~ 10 μm.

-------------------------------------------------------------------------

There are several ways to create megamer (Tomalia, 2005). The megamer should grow in such a way that it would again have a CCU and a surface CNN region, but nano brain seed as the basic unit. If we use different kinds of nano brain seeds with unique dendritic chain inside the megamer (Figure 7 c), or different generations of dendrimers (Figure 7 d), or different kind of molecular switches doped dendrimers, or spherical hollow assembly directly (Figure 7 e) then there are enough freedom to manipulate CCU and CNN properties by tuning the synthesis of the megamer. Here we suggest one alternate method of creating megamer, apart from direct chemical synthesis (Tomalia, 2005). After synthesizing the optimum generation of dendritic architecture or final nano seed with CNN on its top surface, we try to assemble many such seeds using layer-by-layer self-assembly (LBL-ESA, Decher, 1997; Bandyopadhyay, 2003). We use nano brain seeds with negative and positively charged surface (Figure 7 a). For this purpose, CNN cell molecules are anchored with multiple COOH or $NH_2$ groups so that when alternate layers of nano-seeds are grown, heating up the complete architecture at ~ 250-300° C, forms acetamide or –COONH– linkage between any two dendrimers. Thus, a giant single molecular sphere is created. Here $10^3$ cycles (1500 bilayers) of alternate ESA deposition is required, as each bilayer growth saturates in ~20 minutes in solution; it requires automated deposition for 6 days. For fully automated fabrication, we adsorb the nano-brain seed at the top of vertically grown carbon nanotube (CNT) film (Wei, 2002), so that automated deposition produce several nano-brains at a time each placed at the top of a CNT (1.5 nm in diameter). Among several other possibilities, we can also grow micelle architecture (Dou, 2003) using the basic seed dendrimer. Growing a giant H bonded network (Satake, 2005; Huang, 2006) is another choice of growth starting from the basic seed assembly.

# B. Writing a pattern on CNN without destroying CCU configuration and vise versa

The top CNN surface of a megamer or nano brain seed would continuously change the input pattern with time, to derive the solution of a problem written in terms of logic pattern. Note that operational mechanism of nano brain seed and megamer would have significant differences (Figure 7 f). We can follow two distinct protocols to operate the spherical assembly selectively between CNN and CCU. One of them is to write the CNN pattern so that the potential pattern in the CCU is written simultaneously as an effect of same sequence of pulses (Figure 8 a). As CCU and CNN have different relaxation time constant, and respond to different bias magnitudes, particularly designed array of pulses is effective in creating two distinct codes one at CCU other at CNN. Alternately, we can write CCU patterns first, since it requires much higher bias, and then write CNN input pattern, as it requires much lower bias.

In any molecular CNN, encoding an input pattern requires several tricky approaches. As soon as part of a targeted input pattern is written on the surface, evolution of the pattern following CA rules starts instantly. An effective input pattern processed by CNN is very different, as it is changed to a combination of partly evolved and partly targeted input pattern. This is a universal problem and true for any self-assembled CNN cells. One way to confirm that CNN is processing the targeted input pattern effectively is to start writing from a rather different pattern, which would change in course of time in such a way that when final part of the targeted input matrix is encoded, the final pattern at that instant would be the targeted input pattern. One key issue for this method is matching the dynamics of pattern evolution with the writing speed. Computation on the CNN surface is generally uncontrolled and starts spontaneously. By proper choice of input pattern one can initiate spontaneous change in pattern by scanning once at a higher bias. However, to minimize STM influence, spatial gap is created between different parts of the input pattern so that temporal evolution of pattern is controlled accurately.

Note that in the megamer, required electric bias would be much higher for writing CCU and CNN codes, rather than nano-brain seed. However, writing potential pattern in megamer brain would be much easier than the nano brain seed since finite potential barrier between two nano brain seeds in the megamer would prohibit uncontrolled communication among neighbors.

## C. Coherent control of CNN by CCU

A small potential profile change in CCU can change the set of CA rules under operation on the spherical surface. The size of a typical nano-brain seed would be ~ 12-15 nm. The tunneling mechanism is effective around ~ 6 nm separation, therefore a nano-brain would have a quasi-charge (polaron/soliton, Goodson, 2005; Wu, 2006) dominated transport mechanism in the dendritic trees. As the soliton moves with the velocity of sound, CCU to surface CNN communication of a nano-seed takes a few pico-seconds. CCU potential distribution governs transport of polaron waves from the central region to all directions. Simultaneously, change in surface potential distribution during evolution of CNN pattern is influenced by waves propagating through dendritic tree. Coherent control of CCU is modulated by large number of switches doped in the branching trees connected between CCU and the CNN surface neurons in the nano brain seed.

However, since our megamer nano brain size is ~ 10 μm in diameter growth of molecular assembly from 15 nm to 10 μm would die out all codes written in the CCU. Therefore, we developed particular design, translating the CCU code to the final CNN surface. Building such a giant architecture follows particular rules. First, entire architecture is conjugated so that polaron/soliton transport survives throughout the assembly. However, the system may not operate faster than megahertz frequency ($10^6$ Hz). Second, the basic pattern features of the CCU potential map in the space between nano-brain seed and outer CNN surface is created by arranging different nano brain seeds (operates in different bias region in particular geometric shape and pattern. Third, different nano brain seeds are arranged in the architecture so that they do not screen the CNN or any potential distribution created below the surface CNN. Our recent study has shown that if CCU potential distribution is created by arranging more conducting nano brain seeds (operates at very low bias region) in a geometric pattern following $r = a \sin k\theta$, where $r$ is the co-ordinate of doped neuron cluster, $k$ is branch number, $\theta$ is angle of deviation, $a$ is constant particular of a dendritic architecture then the CCU code can CCU strongly influence the CNN surface even at a separation of 20 μm (Figure 8 b). Fourth, since all top layers of the megamer nano brain are covered entirely with nano brain seeds; they create a 3 D cellular neural network of nano brain seeds. However, they follow a new set of CA rules or CNN governing equation.

For a nano brain seed, CNN surface consists of multilevel molecular switches, or molecular neurons, however, for the megamer giant brain, CNN surface consists of nano-brain seeds, each can take thousands of decisions. Therefore, we nearly touch neuron performance in an artificial hardware.

# V. Specific design modifications for nano-brain operating in a nano-factory or inside our body or operating as a supercomputer

## A. Nano-brain operating in human body as a nano-surgeon or nano-doc

Most essential requirement for building a nano-surgeon is to design and synthesize molecular machines operational while floating in the body fluid. Till now, molecular engineers have not adopted any universal solution so that an integrated nano-brain is functional in that solution when coupled to those machines. Every different molecular machine operates in a specialized solution, and there is no guarantee that they will continue to operate if that solution is replaced with body fluid. Once this problem is resolved nano-brain is all set to couple with the existing molecular machines on the CNN surface and would be ready to be injected in a body. However, some molecular machines are driven by electron injection, some of them are pH dependent and some of them even dependent on the ion concentration of the solution (Bertrand, 2000; Schular, 2000). If molecular machines are coupled to the CNN surface, CA operation or pattern based computing is practically stopped. Logic states are written on the molecular machines as instructions and the surface potential distribution is changed as CCU relaxes the system. Nano-brain functioning as surgeon does not require the CNN to compute evolved input pattern.

Inside human body, a nano-brain is considered as foreign element and antibodies attack the assembly destroying its functionality. Therefore, we propose to cover nano-brain with hormones/enzymes known to our body by forming *giant enzyme* complexes with the nano-brain (Perham, 2004). To form *multienzyme complexes*, those particular hormones/enzymes are chosen which specifically act in those targeted place where the nano-surgeon needs to operate. As hormones/enzymes uses lock-and-key trick to reach to a very particular part of our body, nano-surgeon would be transported automatically to that very region. Covering and uncovering of the nano-brain should also depend on pH variation or Na/K ion concentration, so that they are covered with hormones/enzymes once again automatically as soon as antibodies attack them during operation. Covering nano brain with micro-capsules is an alternative method where LBL-ESA deposition technique is used to cover the assembly, and it is released at particular pH similar to a *drug delivery* (Figure 9 a; Sukhorukov, 2001). Finally *non-antibody ligand* could also be a possible choice to anchor with nano brain or it could be captivated inside the ligand-bound nanocarriers (Peer, 2007).

## B. Nano-brain operating in as nano-Factory Control Unit

The supercomputing megamer nano-brain has surface CNN operating particular CA rules. When CNN surface is covered with the maximum possible number of machines, it is nano-doc. In contrast, nano-FCU has to exhibit both kinds of signal processing. To run a nano-factory we need to analyze multiple parallel decisions on the CNN surface. In addition to this, multiple smaller conical/spherical/tubular assemblies covered with large number of molecular machines are attached to the CNN surface to function as working arms of FCU.

Versatile synthesis of dendrites allows fusing two distinct branching of dendrites into a single structure (Galliot, 1997). The primary sphere is the major computing CNN based CA system, and a number of conical/spherical/tubular assemblies covered with large number of molecular machines connected to this primary sphere translate that decision into a well defined sequence of jobs (Figure 9 b; Kurzynski, 2006). Following this particular design, one can analyze and simultaneously execute operations of multiple distinct nano-surgeons together for creating a delicate product with multiple functionalities. Similar design requirement is also demanded by L. Behera's quantum brain model (Behera, 2006). While modeling eye-activities, an implicit requirement of two parallel brains is established. One is a quantum brain, triggering wave-packets to reproduce experimental observations and another one is classical brain processing the eye-sensor data. Here smaller conical/spherical/tubular assemblies covered with large number of molecular machines functions equivalent to the quantum brain, and the nano-brain seed functions as the classical brain. Therefore, nano-FCU may also function as nano-assembler with robotic arms.

## C. Nano-brain operating in massive parallel supercomputer

For using nano-brain as a standalone supercomputer, the spherical assembly has to be covered with a stable spherical cover with thousands of nano-needle electrodes to write input patterns on the surface (Figure 9 c). Nano-needle electrodes must not move relative to each other during operation. The CNN surface should translate reliable output pattern to the electrode arrays.

Supercomputing using the nano-brain means generating differential equation starting from a simple set of pattern (Toffoli, 1984; Biafore, 1994; Margolus, 1984; Wolfram, 1989), which change with time following solution of the differential equation. By changing the CCU potential profile, we instruct the assembly to carry out a particular kind of mathematical operation. Computation does not necessarily means generating pattern that produce say a primary number as a solution of a problem. Rather, the solution would be a generalized pattern reflecting all possible solutions that could arise because of this particular problem. Generalized features are more pronounced than specific features of the problem, for this kind of computation. Mathematical operators and operational tools for the computation process are written in the nature of potential energy distribution in the CCU. Importantly Christopher Davia's brain model (Davia, 2006) also concludes that spatio-temporal pattern of the traveling waves inside our brain is responsible for computation. Solitons, traveling waves and non-dissipative robust waves maintain structure and energy during computation of our brain. However, according to him this condition is valid till they are propagating in the relevant environment. This particular condition enables the system to generate versatile decision-making and global co-operation in biological computation. The CCU potential profile mimics modulation of polaron/soliton length, which is equivalent to Davia's constraint condition.

# V. Conclusion: three key conceptual advancements and a review of consciousness

In this article, we have discussed three key factors to develop our primitive 17 molecular nano-brain to an advanced mm-brain composed of 100 billion neurons. In our previous proto-nano-brain, CCU could not take any independent decision, however, CCU potential profile can now take series of decisions in principle and instruct execution units on the CNN accordingly. Prior to us, CNN models were built using constant resistor, variable resistor or memristor, and rectifier. For the first time, we propose to use multilevel switches to construct CNN. Note that multi-valued logic is essential to construct an adaptable learning hardware (Perkowski, 2002). Our conceptual spherical CNN is apparently first of its kind with distinct advantages over planar CNN. It can be interfaced more efficiently with the outer world; information processing surface area is infinite in principle, computation is more emergent than planar surface as evolution of a pattern essentially considers all bits over the surface equally. Our third and final advancement is in linking a giant CNN to a nano-brain seed using a unique symmetric potential distribution. Note that this is a major paradigm shift from earlier proposed CA based computers (Hillis, 1984).

Two significant experiments, dendritic expression of microtubule assisted protein (MAP2) in rats carried out by Nancy Woolf (Woolf, 2006) and the role of MAP-tau overexpression in the learning and memory of *Drosophila* studied by Mershin *et al* (Mershim, 2006) show that consciousness originates from the very communication between atom determining protein function. At higher scale consciousness is translated from proteins to cells and then from cells to tissue. These formulations are strongly supported by tracking activity of living human brains using *electroencephalography* (ECG) and *magnetic resonance imaging* (MRI). We are studying similar electronic and magnetic responses during nano-brain operation, since redox active atom's activation to dendritic chains and thus communication between CCU and CNN may provide artificial BA map similar to the MRI scan of the real brain. The Hameroff's argument (Hameroff, 2006), that consciousness is secondary response of a metastable pattern generated by synaptic pulses has close similarity to our studies. The correlation between BA dynamics for CCU and CNN maps the metastable state and the conscious output. Alternate to this spherical nano-brain, a helical assembly could also be constituted (Kornyshev, 2007), since microtubules may play major role in neuro-cognition and permanent memory in our brain (Woolf, 2006).

---

**Acknowledgements**: Authors acknowledge Dr. John Liebeschuetz, The Cambridge Crystallographic Data Center, 12 Union Road, Cambridge, CB2 1EZ, UK for providing dendrimer structures and critical review of the work. Authors also acknowledge Prof. Michael J Cook from UEA, Dr. Jonathan Hill and Dr. Y. Wakayama from NIMS for their contribution in several parts of the results presented here. This work is part of the Japanese Patent filed JP-2006-19552.

**Competing interest statement** The authors declare that they have no competing financial interest.

**Correspondence** and requests for materials should be addressed to A. B. (anirban.bandyo@rediffmail.com)

Hess, H., Bachand, G. D., Vogel, V. (2004) Powering nanodevices with biomolecular motors *Chemistry - A European Journal, 10*, 2110-2116.
Hall, J. S., (2007) Beyond AI: Creating the Conscience of the Machine, Prometheus Books, pp-408.
Hawker, C. J., Wooley, K. L., (2005) The Convergence of Synthetic Organic and Polymer Chemistries *Science, 309*, 1200-1205
Hipps, K. W., Hoagland, J. J., (1991) Top Metal and Bias Effects in the Tunneling Spectrum of Copper(II) Phthalocyanine. *Langmuir. 7*, 2180-2186.
Huang, W. Zhu, H.-B., Gou, S.-H., (2006) Self-assembly directed by dinuclear zinc(II) macrocyclic species *Coord. Chem. Rev. 250*, 414-423.
Hillis, W. D., (1984) The connection machine: A computer architecture based on Cellular Automata, *Physica D: Nonlinear Phenomena, 10*, 213-228.
Hameroff, S., (2006) Conciousness, neurobiology and quantum mechanics: The case for a connection, pp-193-242, *The emerging physics of consciousness,* Tuzsynski, J. A. (Ed.), Springer Berlin Heidelberg, 2006, printed in Germany.
Husband, C.P.; Husband, S.M.; Daniels, J.S.; Tour, J.M. (2003) Logic and memory with nanocell circuits, *IEEE Transactions on Electron Devices*, 50, 1865 – 1875.

Jian, H., Tour, J. M., (2003) En Route to Surface-Bound Electric Field-Driven Molecular Motors *Journal of Organic Chemistry* 68, 5091-5103.

Koumura, N., Zijlstra, R. W. J., van Delden, R. A., Harada, N., Feringa, B. L. (1999). Light-driven monodirectional molecular rotor *Nature 401*, 152-155.
Kohonen, T., (1989) Self-organisation and associative memory, 3$^{rd}$ edition, Springer Verlag Berlin Heidelberg, pp-312, printed in USA
Khondaker S. I., Yao, Z., (2002) Fabrication of nanometer spaced electrodes using gold nanoparticles, *Appl. Phys. Lett. 81*, 4613-4615.
Kurzynski, M., (2006); Structure of biomolecules and dynamics of biomolecules, *The thermodynamic machinery of life* Springer Berlin Heidelberg, printed in Germany.
Kornyshev, A. A., Lee, D. J., Leikin, S., Wynveen, A., (2007) Structure and interaction of biological helices, *Rev. Mod. Phys. 79*, 943-996.

Li, C., Fan, W., Lei, B., Zhang, D., Han, S., Tang, T., Liu, X., Liu, Z., Asano, S., Meyyappan, M., Han, J., Zhow, C. (2004) Multilevel memory based on molecular devices *Appl. Phys. Lett. 84,* 1949-1951.
Liang, W., Shores, M. P., Bockrath, M., Long, J. R. & Park, H. (2002) Kondo resonance in a single-molecule transistor. *Nature 417*, 725–729.

Metzger, R. M., (2000) Unimolecular rectification down to 105 K and spectroscopy of hexadecylquinolinium tricyanoquinodimethanide Synthetic Metals, 109, 23-28.
McGinness, J. E., Corry, P., Proctor, P. H., (1974) Amorphous semiconductor switching in melanins, *Science 183,* 853-855.
Mukherjee, B., Pal, A. J., (2005) Dielectric properties of (multilevel) switching devices based on ultrathin organic films, *Chem. Phys. Lett. 401*, 410-413.

**Figure Captions**

Figure 1. **Fundamentals of Major conceptual shift: a** Molecular monolayer is represented as 2 D array of CNN cells. Noise function for the vertical measurement is $\Psi(t)$ and for horizontal measurement through an organic monolayer, it is the product of all junctions in the probabilistic path each contributing as $\Omega(t)$ (top). Nano-cell project has addressed horizontal noise in an innovative way (see Husband, 2002, Dutta, 1981). A Gaussian expanding noise is plotted at A, B, C points for three different rates of expansions, the triangular geometry does not change, and therefore decision remains the same. For vertical measurements, surface propagating local noise have minimal effect on the final decision. **b** Schematic presentation of an conventional hardware is denoted as A and an adaptable hardware is denoted as B. A is composed of binary switch, therefore if there is a permanent change on the processing circuit, the computation is lost, i.e. it cannot adapt. B is composed of multilevel switches therefore, even after permanent changes; there are possibilities (top). Line profile of potential for an adaptable hardware is measured as an average spectrum (below, left). Four examples of potential profiles, -bold plotted pattern denoted as C depicts the line profile marked in A, or background potential. The dotted profile crosses this line, therefore is not allowed. **c** Potential profile of a nano-brain in the configuration space, undergoing self-organization, arrow denotes an input logic pattern, A, B, C are self-organization, D is solution of a problem. **d**. Schematic presentation of transition from existing concept of linear communication to radial communication in a disk shaped object, and then in a sphere (left). The schematic presentation of functional parts of a 3D nano-brain hardware, it could be a nano brain seed of ~15 nm or a megamer of ~ 20 μm. Here, A is CCU, B is the boundary of brain seed, C is the symmetry transformation region, and D is the CNN region (right).

Figure 2 **Molecular neurons, design and realization**: **a** Four reversible electronic states of DRQ molecule. **b** Cyclic voltammetry and phase transition relationship, only those states differed by more than 24 mV are distinct in room temperature. **c**. Schematic presentation of step function (top), an integrated spectrum of step function is identical to Kondo response (below). 2T is the total width of bias variation where a ZBC is observed. **d** Current voltage characteristics of a Fluorocene molecule that shows spectrum identical to the Rose Bengal molecule. Its conductance spectrum (top left) and the height profile of Rose Bengal on Au (111) at different bias are shown (top right). Single molecule STM image (below), molecule is 1.1 nm long and 1.3 nm wide. **e** Conceptual memristor's equivalent circuit.

Figure 3. **Cellular Neural Network (CNN), or CA monolayer fabrication: a** Molecular monolayer is first converted into a matrix of binary trinary or tetanary bits and then between two matrices basic changes at the smallest clusters of cells are determined, these basic rules are CA rules of the monolayer. **b** STM image of negatively changed DDQ (-1) monolayer scale bar is 20 nm, scanned at 0.05 nA tip current and 0.88 V tip bias. A cluster of molecules change its direction by a particular angle, so we define one orientation as P and the other as Q. Elementary changes between P and Q is the CA rule. **c** STM image of the organic monolayer of Rose Bengal molecule at four particular biases for a fixed tip current 0.05 nA. One molecule is 1 nm in length and 0.32 nm in width inside the monolayer. **d** Rose Bengal single molecule device between Au (111) electrodes (left). Molecular model of a Rose Bengal monolayer, on a mono-atomic linear chain surface.

Figure 4. **Dendritic assembly design and its architecture: a** The architecture of a PAMAM dendrimer. Shaded regions within the dotted area are doped multilevel molecular switches or molecular neurons. **b** Examples of generation 2 (G2) PAMAM dendrimer with molecular switches on its surface, the surface is covered with DRQ 2 bit molecular switches, architecture is energy minimized using density functional computation. **c** Hydrogen bonding and van-der Waals surface plot of the dendrimer covered with molecular neuron. **d** Quadrupole moment contour image of a material. Dark shade denotes negative and lighter shade denotes positive charge region. **e** STM image of G 6 PAMAM balls on an Au(111) surface, most of the balls are assembled at the step edges, image is taken at 0.91 V bias and 0.05 nA tip current. **f** If the dendritic architecture has only one symmetry then CV appears as the left spectrum. All redox active sites accept electrons at once at a single bias, and releases all at some other bias. Two distinct structural changes are visible. If the dendrimer has multiple distinct symmetries then equivalently number of redox peak increases, for every single peak a distinct conformer exists. Note that the dendrimer may take more than one electron corresponding to a peak.

Figure 5. **Design and analysis of Central Control Unit: a.** Basic molecular architecture of the dendrimer core and schematic presentation of the potential density distribution in the CCU. Five molecular neuron examples A, B, C, D, E are presented here. Depending on the molecular conducting state; the CCU core changes its potential. The schematic presentation is based on theoretical simulation of five molecules, DDQ (A), Rose Bengal (B), DNA (C, only one periodic unit is used), OH/NH$_2$ doped C$_{60}$ (D) and phthalocyanine (E). **b** Variation in polaron transport

length L with angle θ, L is determined by measuring the distance between negative potential boundary surface of CCU and the outer CNN surface. In practice, polaron transport length does not follow this principle accurately but this particular scheme provides a consistent output. We fix a neuron molecule or CNN cell on the spherical surface, and angle that it makes with another CNN cell on the surface with respect to the center is the angle θ. L variation for two particular CCU potential cases are plotted here, each CCU potential generates a particular set of rule, say set A (having *n* distinct rules), set B (having another *n* distinct rules, *n* is a natural number). **c** If a point charge moves through the surface of the CCU, the potential measured is a function of θ, CCU(θ). Similarly we get another potential function CNN(θ), and these two potentials are correlated with a function of L, f(L), which measures CCU initiation strength in determining the CNN rules. **d** Pristine role of CCU is extracted from the mixed response of a nano brain using basin of attraction (BA) method. Each point on BA plot is a m×n logic matrix.

Figure 6. **Translation operation between CCU and CNN: a** The schematic potential variation of a nano-brain (left) and the potential of CCU (right) with the generalized time co-ordinate. The potential variation of CCU, appears as staircase compared to CNN potential variation. Fluctuation in input pattern on the CNN surface does not vary more than $\Delta E$, CNN input pattern is so selected that it reaches equilibrium at $< \Delta t$, each step of staircase correspond to particular potential distribution of CCU, so that till computation finish, the dendrimer core does not change. **b** Demonstration of spontaneous relaxation dynamics of a $(RB/PAH)_{15}$ supramolecular assembly, packed in a ITO and Al coated sandwiched device, a possible staircase like degradation is observed. Capacitance variation in a continuous bias loop shows threshold charge storage, permanently. This phenomenon may prohibit continuous use of nano brain (it will be tired). **c** The structure of potential distribution in the CCU determines $\Delta E$ and $\Delta t$. If any section of potential distribution in the CCU is symmetric, and built with many such parts (left), then $\Delta E$ is lower and $\Delta t$ is higher, however, if largely asymmetric, say connected by very small region between two large potential spheres (right end), then $\Delta E$ goes higher and $\Delta t$ goes lower.

Figure 7. **Translation region construction connecting CCU and CNN: a** To construct large assembly of dendrimers or megamers, each dendrimer nano seeds are to modified such that some of them are negatively charged with end group $-COO^-$, and some of them are positively charged with $NH_3^+$. **b** Construction of a megamer nano brain. Dendrimers with particularly designed cores are arranged very uniquely at the central region to construct the CCU, and similarly, another kind of dendrimer that generates large number of distinct conducting states, processes RAM and ROM

operations are assembled on the top of the megamer, to construct the CNN. Different layers of concentric spherical electrode assemblies' cover 50 μm diameter nano-brain. **c** Modulating the electronic transport through the dendritic chain is possible by doping redox active groups at various possible locations in the chain, the very central region could also be changed by replacing the core with different molecules or functional groups. **d** The complete megamer architecture could be built using dendrimers of different generations eventually leading to a complete sphere. **e**. The simplest nano-brain architecture could also be constructed on a microsphere, made by metallic or semiconducting nanoparticles, or even surfactants/designer molecules. **f** A table comparing nano brain seed and a megamer brain is presented.

Figure 8. **Symmetry and potential of translation region: a** The basic pulse array sent to a nano brain seed/megamer to test memory and information processing ability of a device (left, top). If the dendrimer responds to the input pulse positively, both read current pulses would differ (left, bottom). The measurement circuit used for this purpose has a function generator and an oscilloscope (middle) attached to the STM. To detect all processable states, a continuous array of pulses with varying magnitude is applied (right). **b** Different possible symmetry structures that coherently connect CNN and the CCU. Co-ordinates of the highly conducting dendrimers, (highly doped with conductance switches, and denoted as shaded balls) are located following an equation $r = a \sin k\theta$. If the periodicity constant $a$ varies, then we derive significantly different arrangements of dendrimers. Energetically allowed values of $k$ for these arrangements fall into Fibonacci/Hemchandra series (top). If $a$ is constant, then solutions fall into the L. Grandi's lotus/rose series (bottom).

Figure 9. **Design modifications for medical application, factory building and supercomputing: a** Design of a nano-surgeon, surface covered with drugs or operating machines, and camouflage of nano-surgeon is done by covering with enzyme, microcapsules and pH dependent materials **b** Design of a nano-factory control unit, particular regions covered with machines interface with products, these interface region are potentially most active region of the CNN surface. **c** For the nano brain to function as a supercomputer, a particular region is selected for the measuring output (cone shaped electrodes), for the nano brain seed, interfacing with multiple electrodes is nearly impossible, however, in a megamer brain one may attach one electrode in every ~150 nm$^2$ area.

Figure 1.

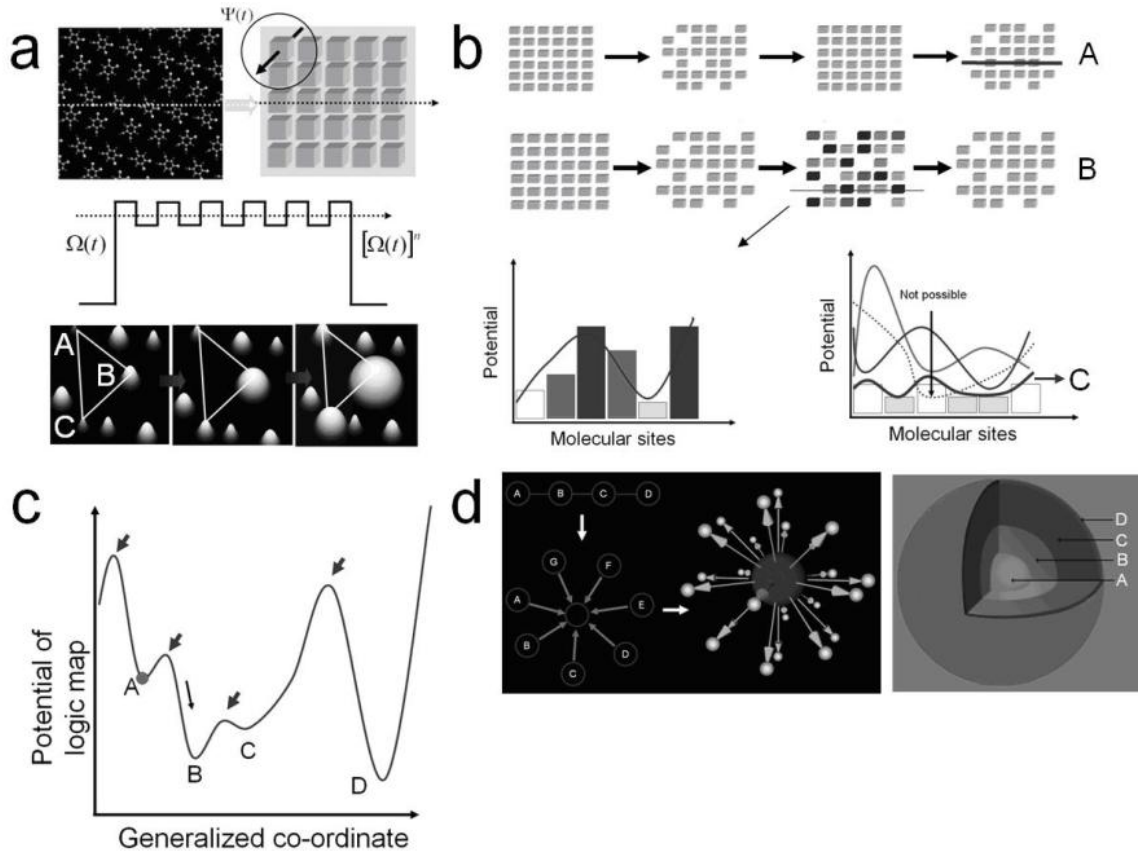

Figure 2

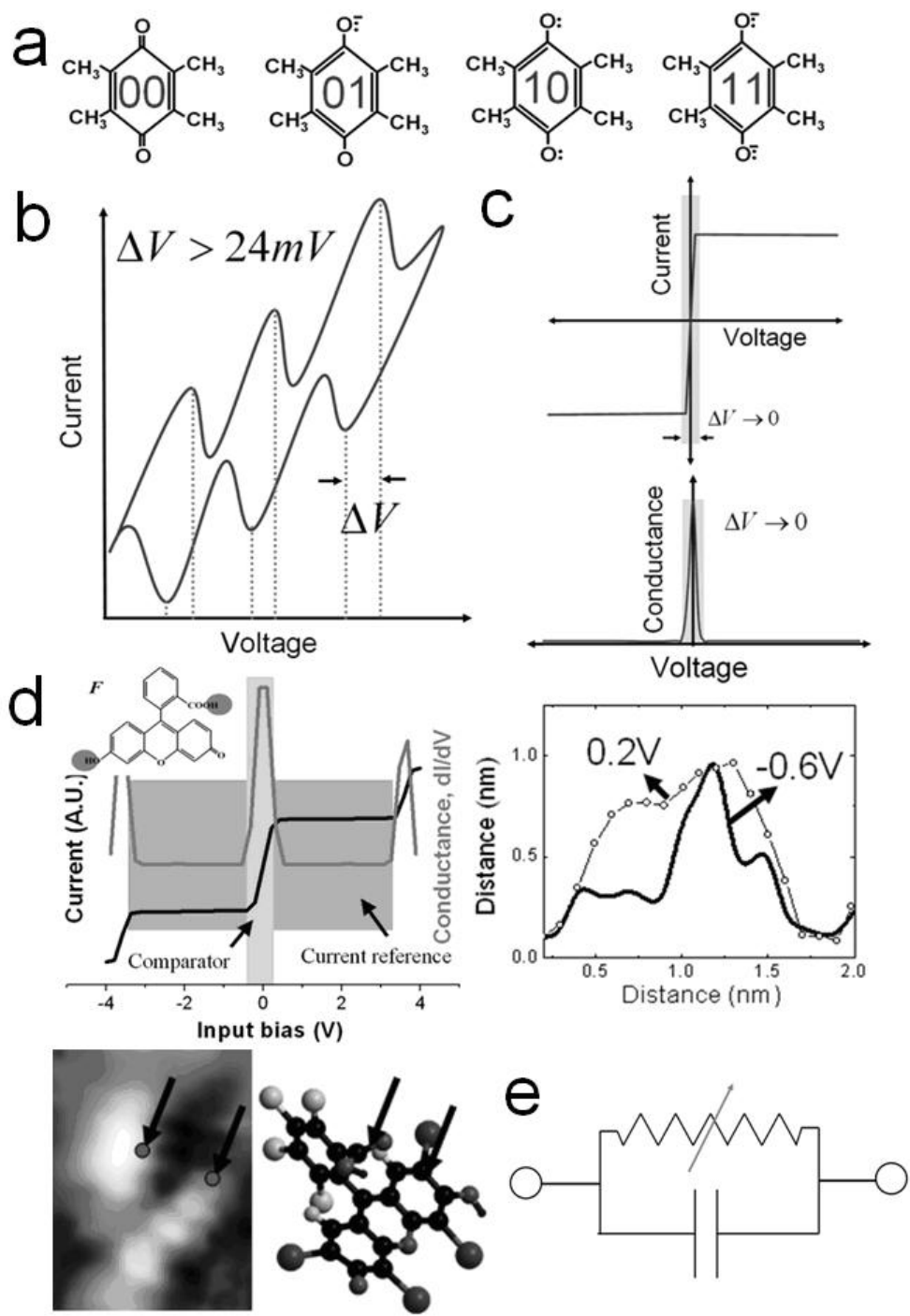

Figure 3.

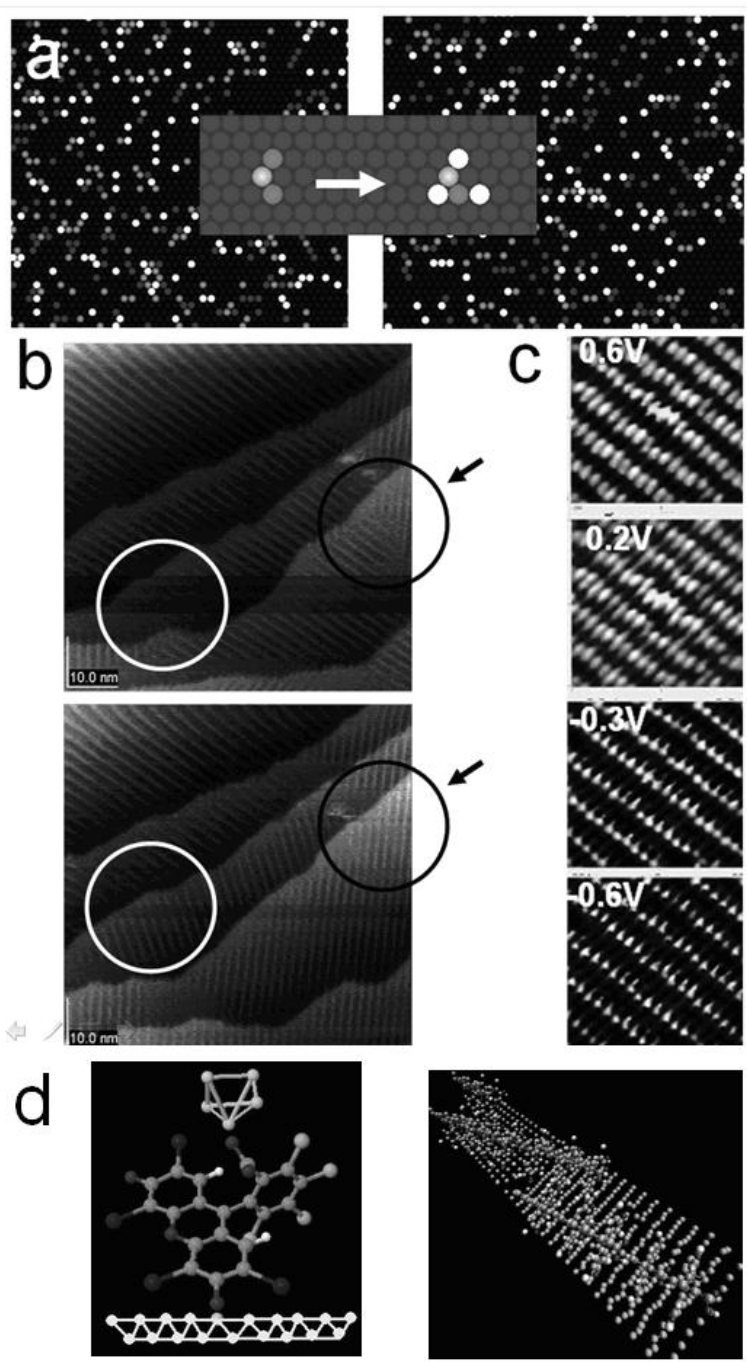

Figure 4.

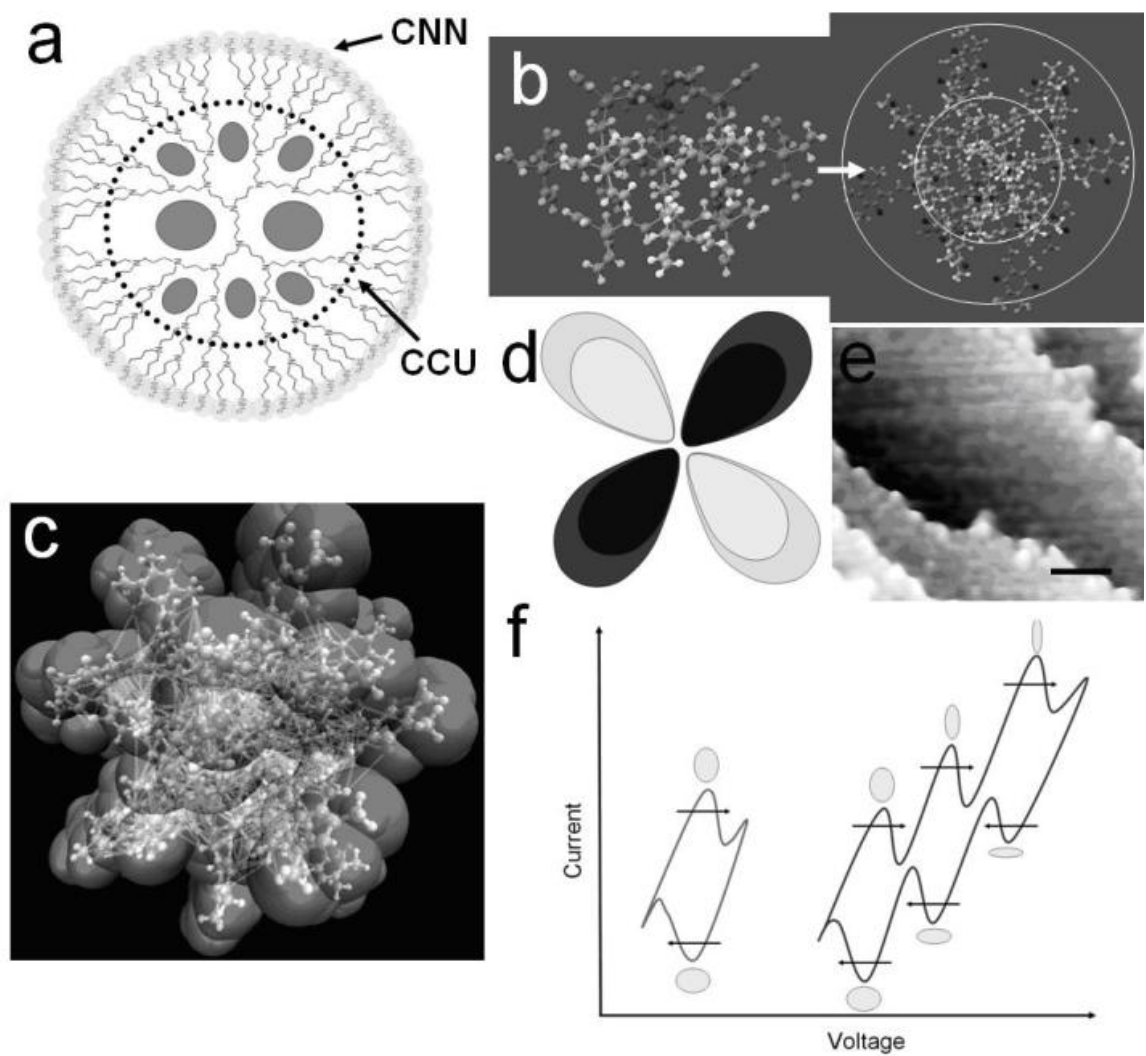

Figure 5.

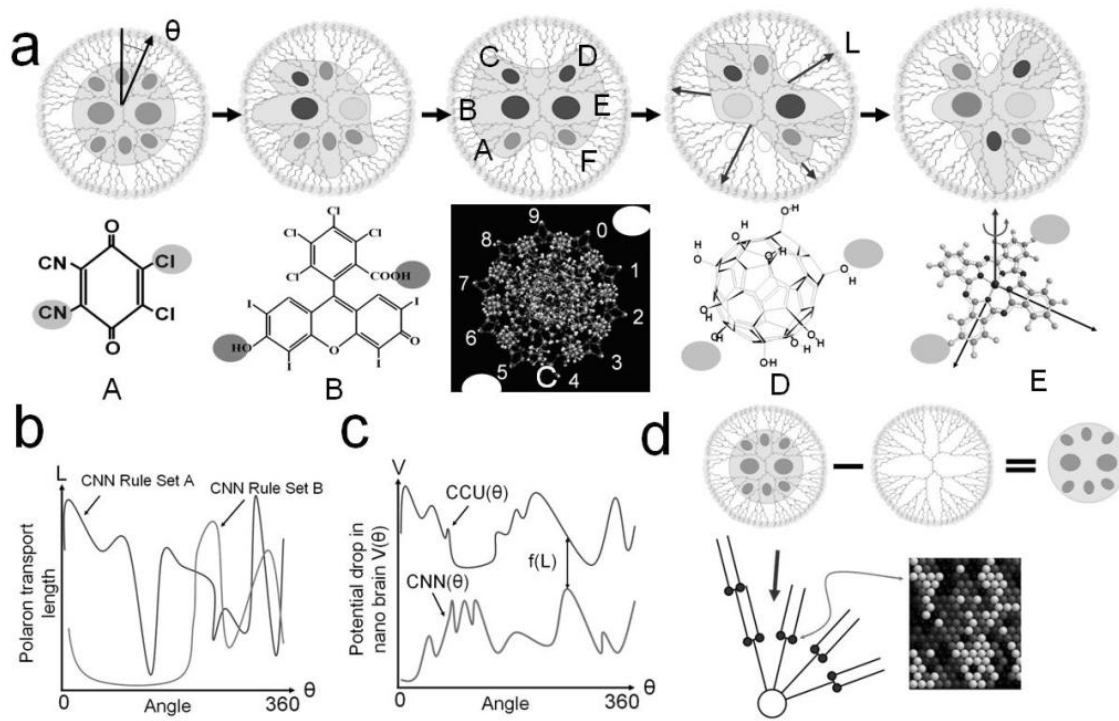

Figure 6

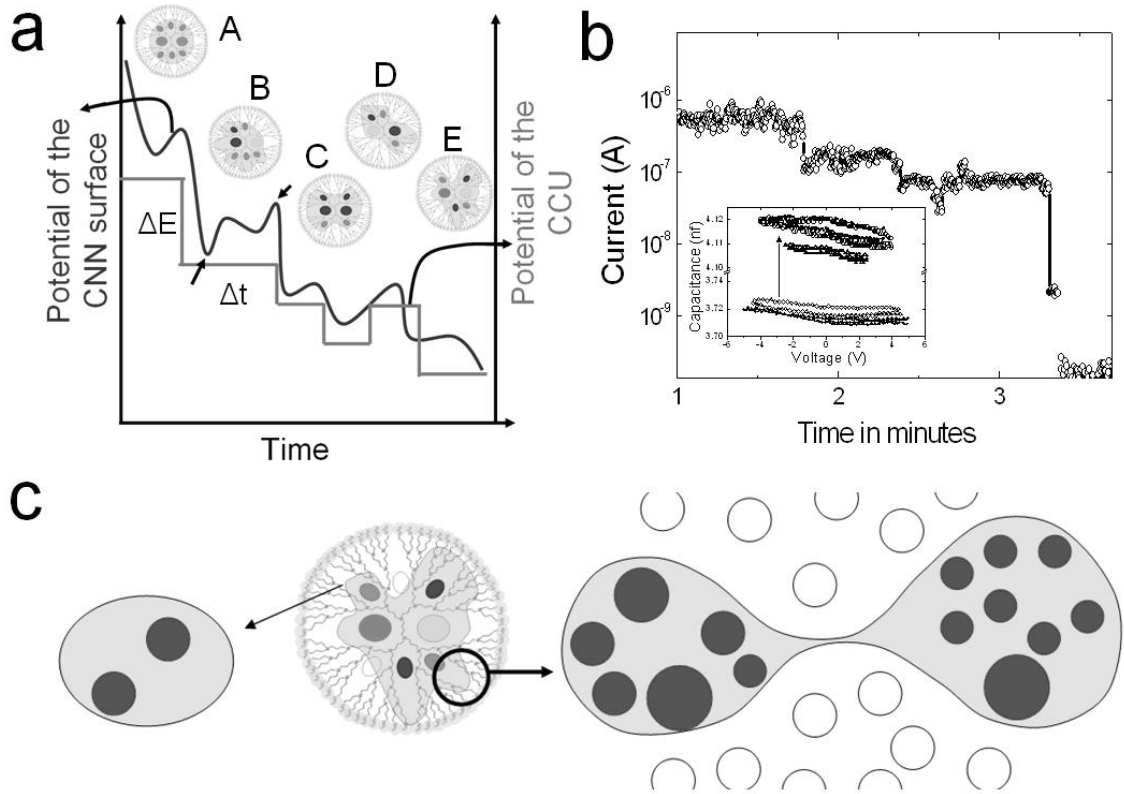

Figure 7

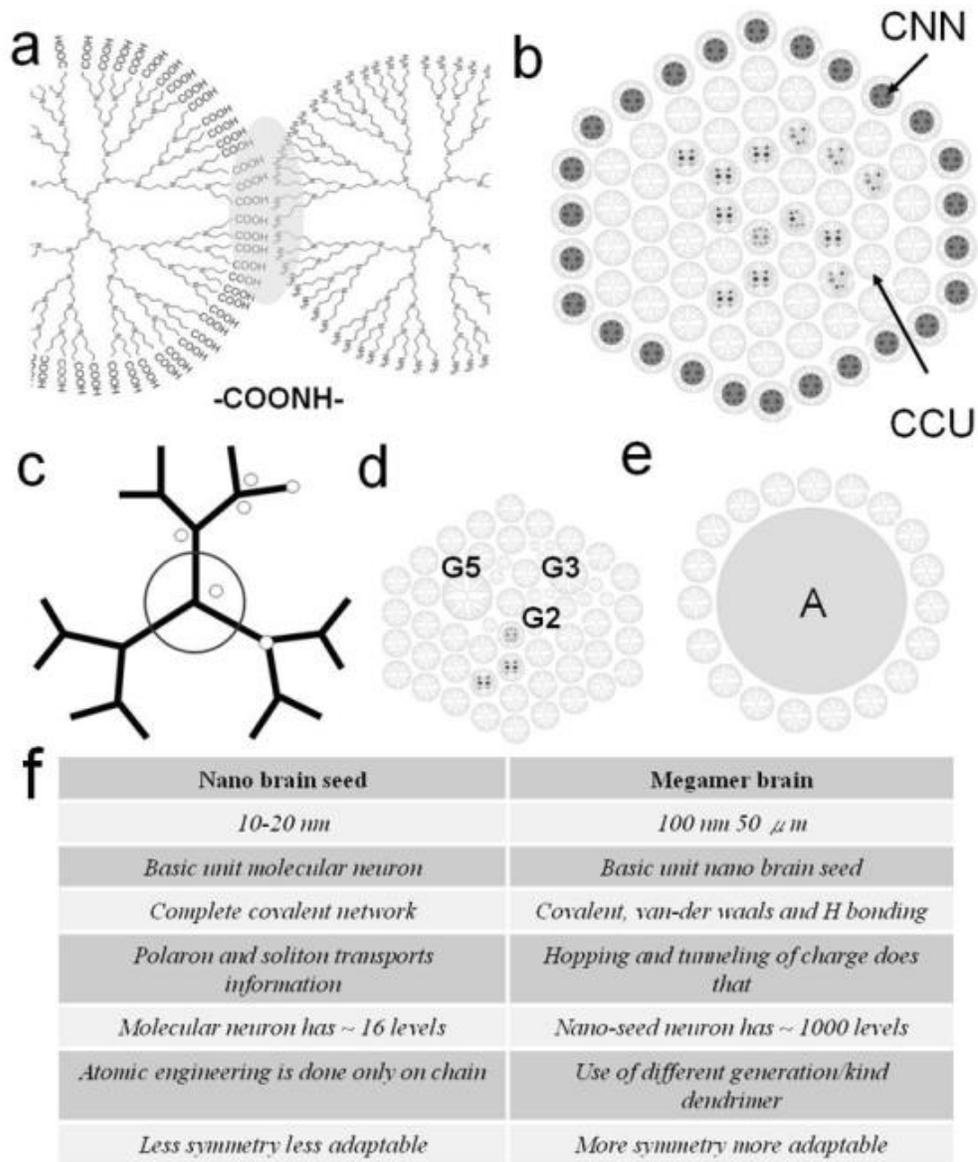

Figure 8.

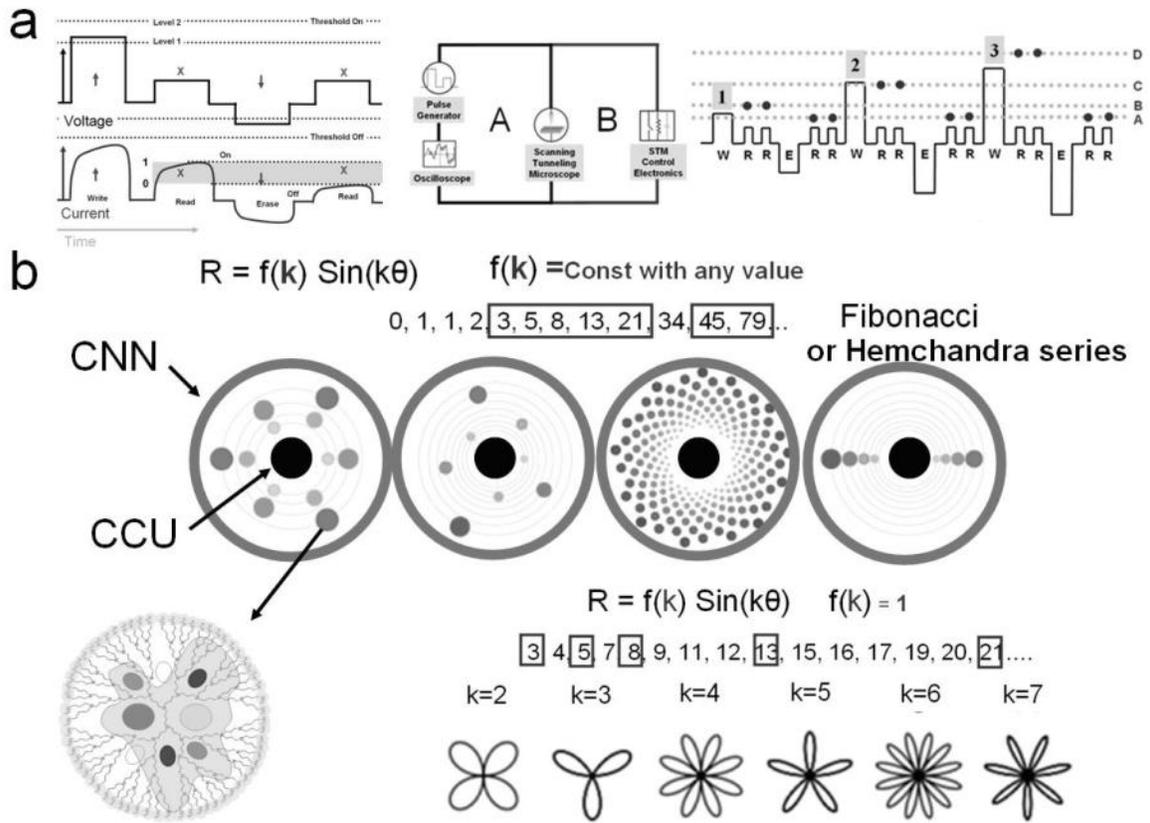

Figure 9.

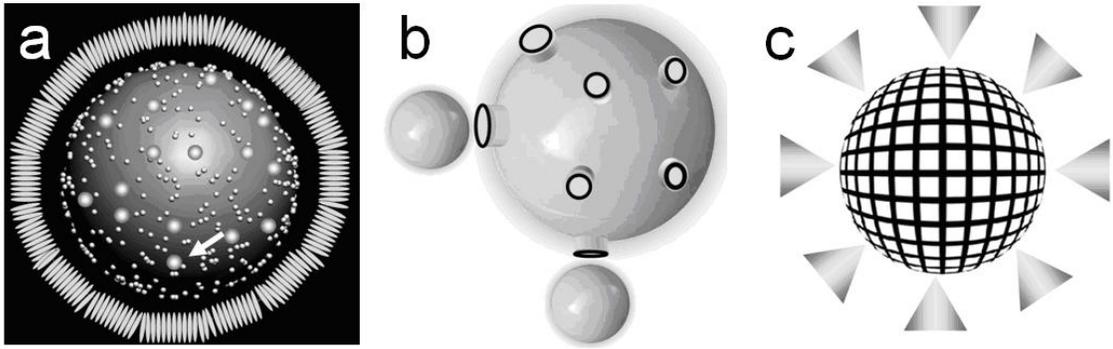